\documentclass[11pt,twoside]{article}

\usepackage{asp2004}
\usepackage{epsf,psfig,graphicx}
\usepackage{lscape}
\usepackage{natbib}

\begin{document}
\title{Small-scale dissipative structures of diffuse
  ISM turbulence: I -- CO diagnostics}
\author{Hily-Blant P., Pety J.}
\affil{IRAM, 300 rue de la Piscine F-38406 Saint-Martin
  d'H\`eres}
\author{Falgarone E.}
\affil{\'Ecole normale sup\'erieure \& Observatoire de
Paris, LERMA/LRA, 24 rue Lhomond, F-75231 Paris Cedex 5}

\begin{abstract}
  Observations of translucent molecular gas in $^{12}$CO and
  $^{13}$CO emission lines, at high spectral and spatial resolutions,
  evidence different kinds of structures at small scales: (1)
  optically thin $^{12}$CO emission, (2) optically thick $^{12}$CO
  emission, visible in $^{13}$CO(1-0), and (3) regions of largest
  velocity shear in the field, found from a statistical analysis. They
  are all elongated with high aspect ratio, preferentially aligned
  with the plane-of-the-sky projection of the magnetic fields. The
  latter structures coincide with the former, shown to trace gas
  warmer and more diluted than average.  Combining our data to
  large-scale observations of poorer spatial resolution, we show that
  the regions of largest velocity shear remain coherent over more than
  a parsec. These filaments are proposed to be the sites of the
  intermittent dissipation of turbulence.
\end{abstract}

\section{Introduction}
Star formation proceeds by condensating gas. While the steps
from dense cores to hydrogen burning stars are basically
identified, what triggers the building of these condensation
seeds remains largely unknown.

Turbulence in the cold neutral medium is driven at large
scale by the warm neutral flow. Dense cores thus form in a
turbulent molecular gas with typical linewidth
$3-4$~km~s$^{-1}$.  The role of magnetic fields during their
formation is still a matter of strong debate
\citep{crutcher1999,padoan2001}. Observations of molecular
lines show that the condensation proceeds at the expense of
the turbulent support, as traced by the linewidth, and that
some turbulent dissipation must take place in the molecular
gas. Turbulence is dissipated at small scales by
viscosity. But how small is this dissipation scale~?
Small-scale structures have been observed in the ionized and
atomic gas down to few AUs, direct observations of molecular
small scale structures are less common. Direct evidence of
small-scale molecular structures include \cite{pan2001}
($10^4$~AU with density 1000 to 5000 cm$^{-3}$) and
\cite{heithausen2002,heithausen2006} (few hundreds AU and
densities $>2\times10^5$~cm$^{-3}$, this volume). Indirect
evidence of such small scale structures (200~AU) were 
inferred from translucent gas observations by
\cite{falgarone1996,falgarone1998kp}.

\begin{figure}
  \begin{center}
	\includegraphics[width=0.6\textwidth]{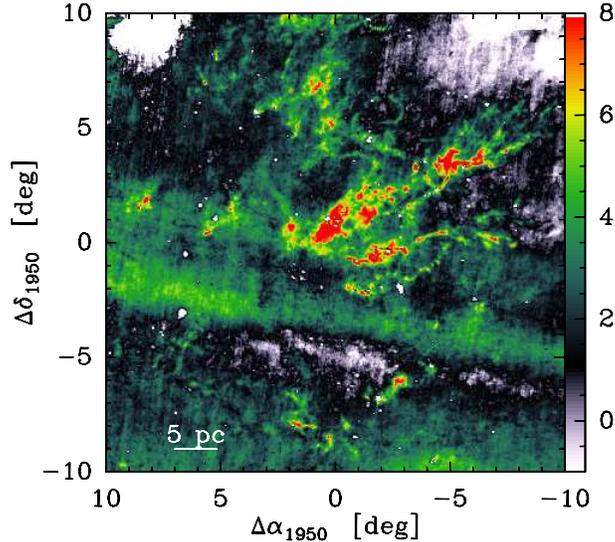}
	\caption{100$\mu$m emission of the cold dust (in MJy/sr)
	  towards the Taurus molecular cloud, calculated using
	  the reprocessed IRAS data \citep{miville2005}. Stripes
	  are due to zodiacal light.}
	\label{fig:iras}
  \end{center}
\end{figure}

Fig.~\ref{fig:iras} shows the 100\micron\ emission of cold
dust in the Taurus molecular cloud, computed following
\cite{abergel1994}. Many small filaments, often associated
with known Lynds objects, are seen. Our work focusses on
regions in the vicinity of these filaments, characterized by
a low visual extinction, ($A_V\le1$~mag) and 100\micron\
intensities $\la7$~MJy/sr, to disclose their formation
mechanism, and their link to dense core
formation. Our large maps, with high spectral resolution
($\approx 5\times10^6$ or 0.055~km~s$^{-1}$ at 115~GHz),
extended the dataset of \cite{falgarone1998kp} and allow to
perform statistical analysis of the velocity field to relate
the small-scale structures to turbulence.

\section{Diffuse molecular gas at small scales}

Fig.~\ref{fig:maps} (top) shows maps of the integrated
$^{12}$CO and $^{13}$CO(1-0) emissions towards the
Polaris Flare \citep{falgarone1998kp,hilyblant2006ii}, done
with the IRAM-30m telescope. The $^{13}$CO emission is
distributed along a main structure with a position angle
P.A.$\approx110$ degrees. This structure is connected to a
dense core, visible at the eastern edge of the map
\citep{gerin1997,heithausen2002}. There are 
areas and velocity intervals in the field where
only $^{12}$CO is detected
and the $^{12}$CO over $^{13}$CO(1-0)
line ratio is greater than 30.  These intervals are called
line-wings. The following work is devoted mainly to the
physical properties, kinematics and dynamics of the emission
in these line-wings.

\subsection{Small-scale optically thin $^{12}$CO filaments}

\begin{figure}
  \begin{center}
	\includegraphics[width=0.28\textwidth,angle=-90]{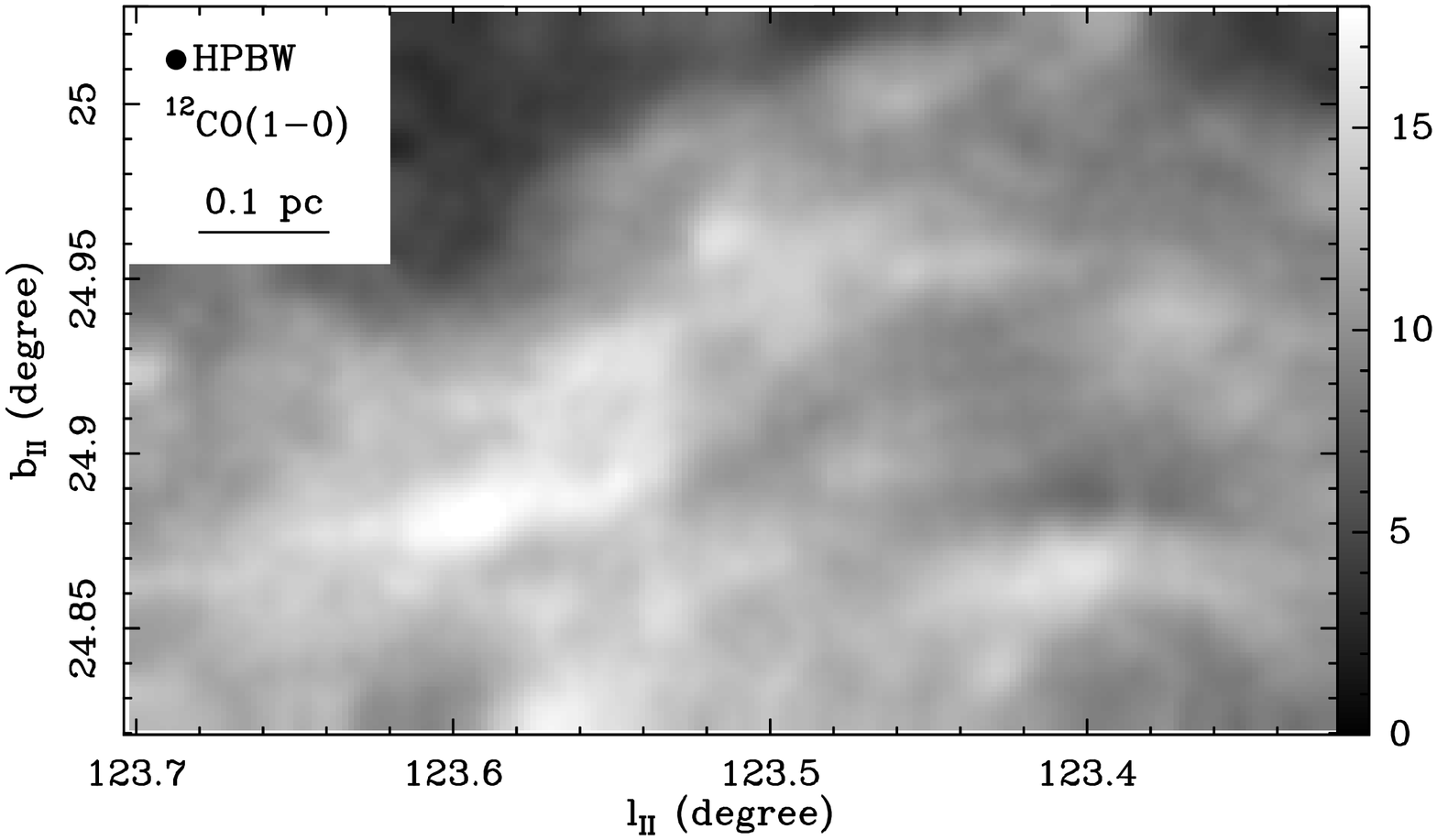}\hfill%
	\includegraphics[width=0.282\textwidth,angle=-90]{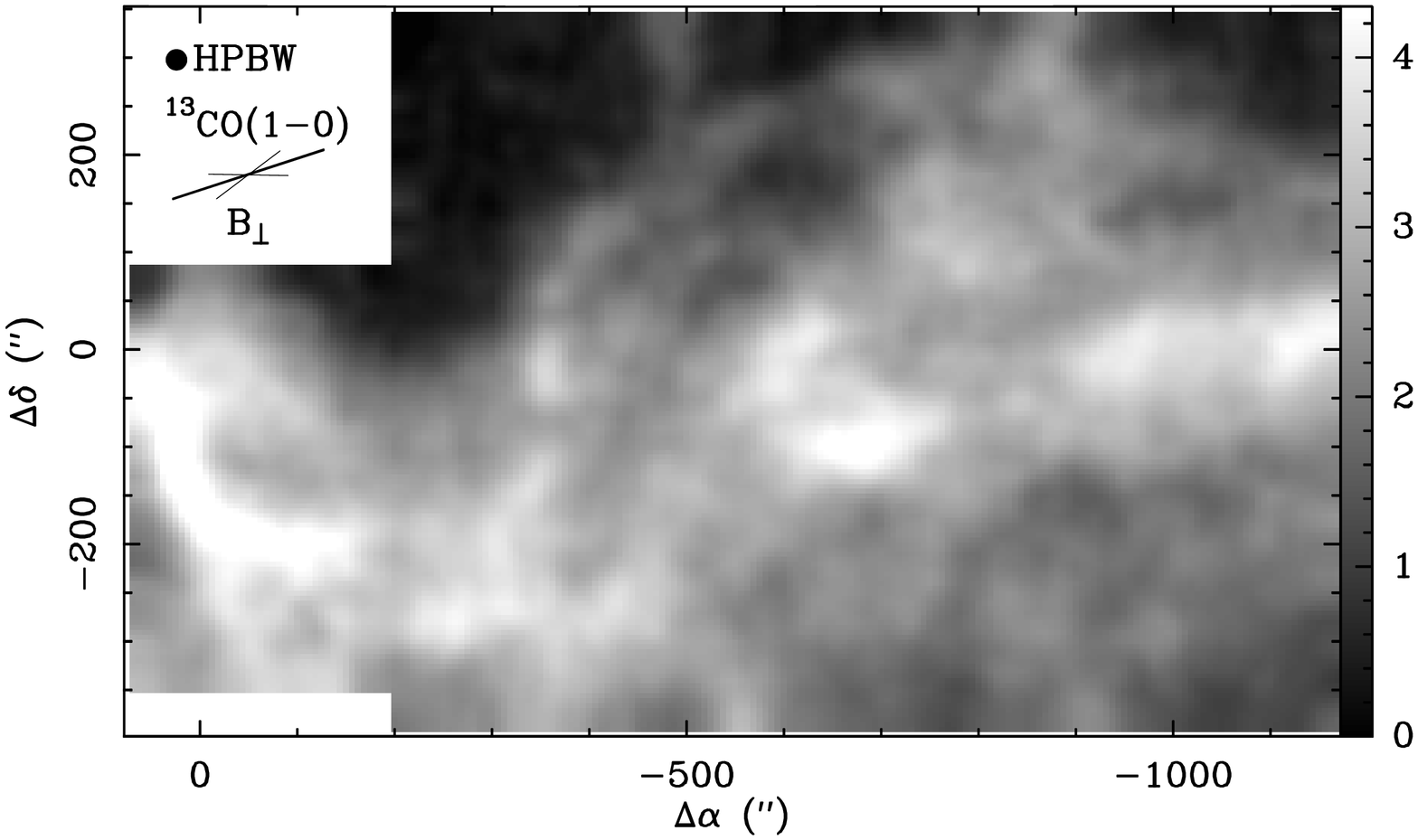}\\
	\includegraphics[width=0.47\textwidth]{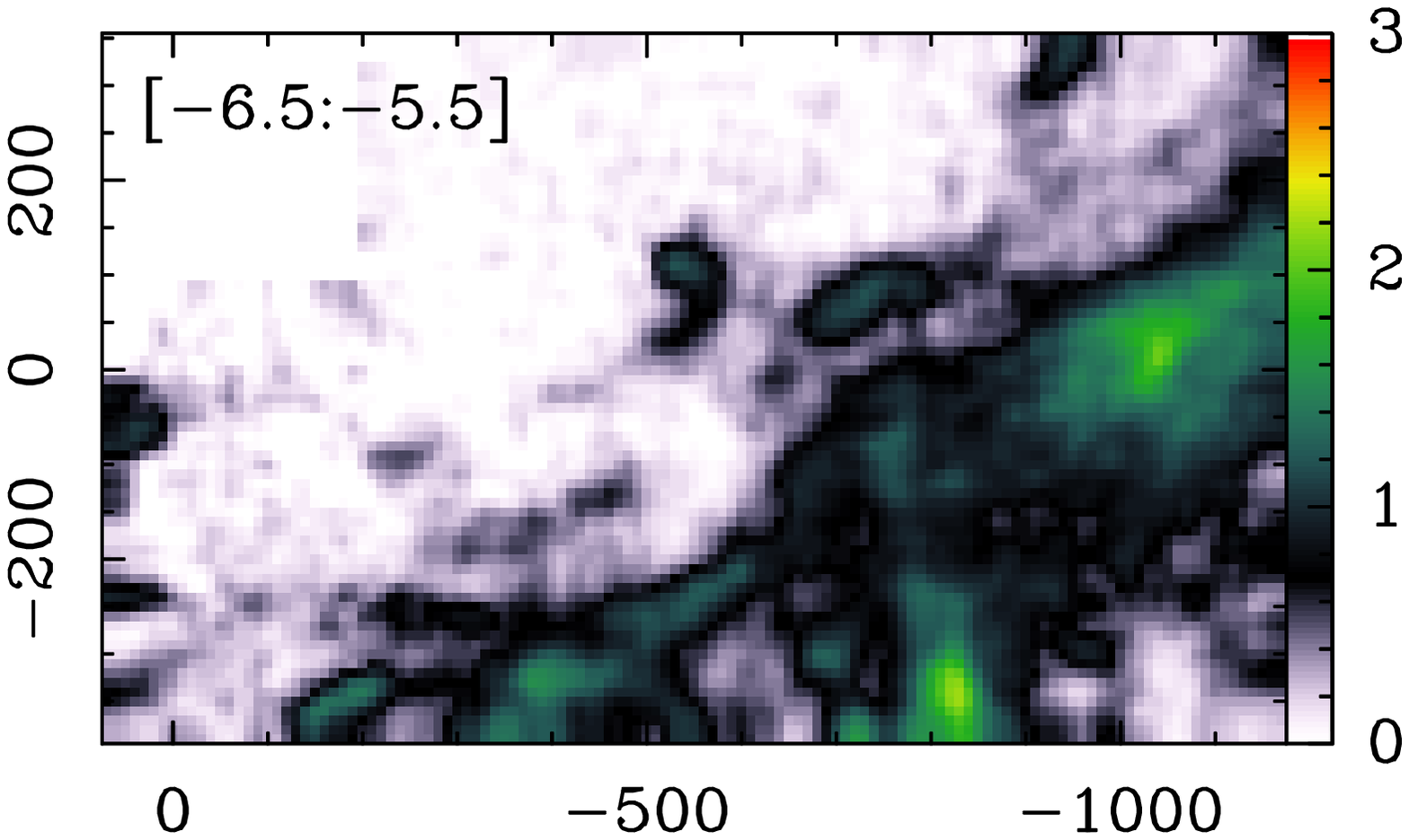}\hfill%
	\includegraphics[width=0.47\textwidth]{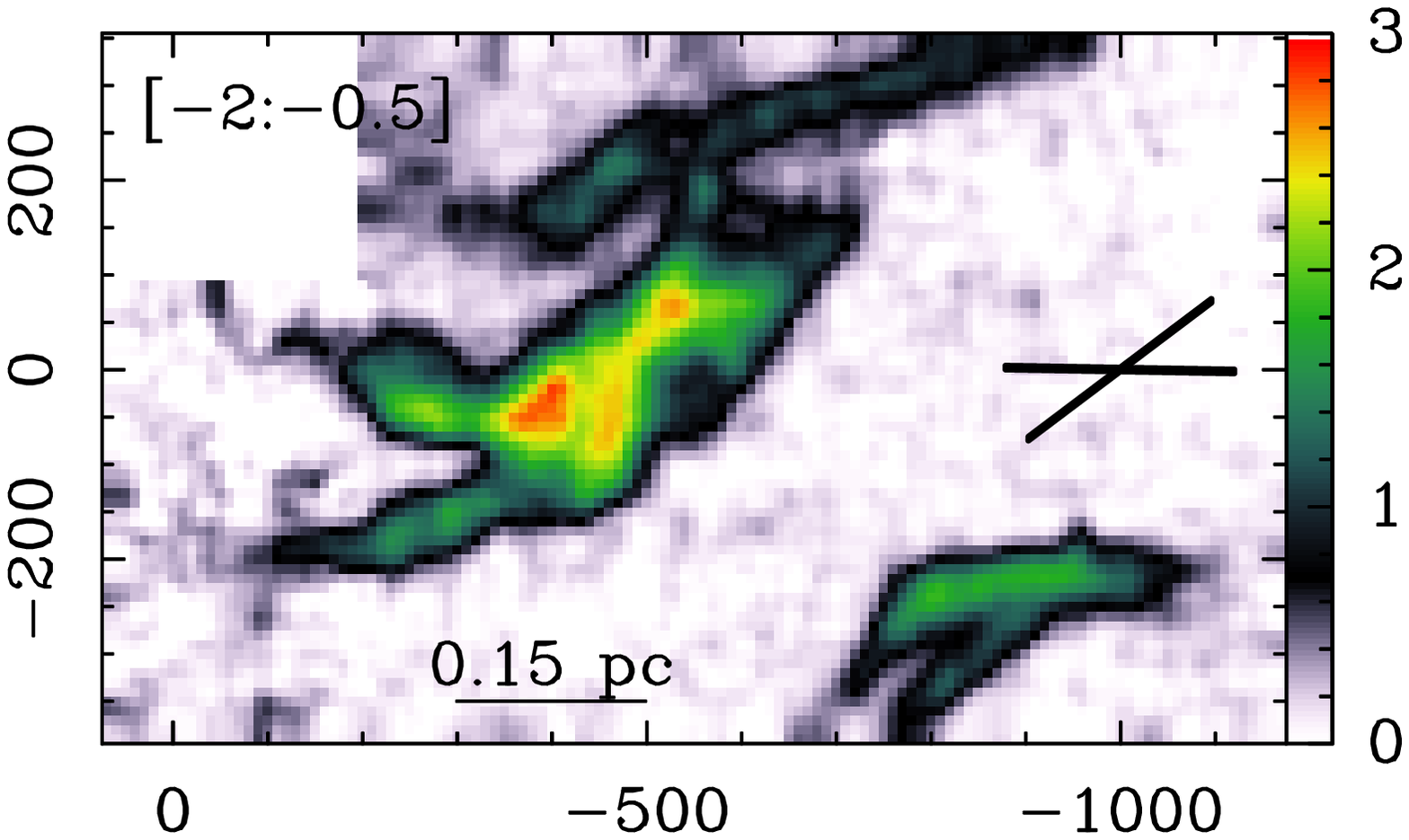}
	\caption{\emph{Top}: integrated emissions of the
	  $^{12}$CO (left) and $^{13}$CO(1-0) (right)
	  transitions. \emph{Bottom}: $^{12}$CO emission in the
	  linewings intervals (see text): $[-6.5:-5.5]$ and
	  $[-2:-0.5]$~km~s$^{-1}$. The direction of the magnetic
	  fields lies within the $1\sigma$ uncertainties
	  as indicated.}
	\label{fig:maps}
  \end{center}
\end{figure}

Fig.\ref{fig:maps} (bottom) shows the $^{12}$CO(1-0)
emission in the wing intervals ($[-6.5:-5.5]$ and
$[-2:-0.5]$~km~s$^{-1}$). Small elongated structures with
typical transverse size 0.04~pc and velocity width
0.35~km~s$^{-1}$ are seen, and with aspect ratio as high as
10. While the upper structure ($\delta b>200\arcsec$) is
also seen in $^{13}$CO at other velocities, all the others
are not detected in any channel in $^{13}$CO. LVG analysis
using $^{12}$CO(1-0) and (2-1), combined with an upper limit
on the H$_2$ column density provided by the visual
extinction \citep{cambresy1999}, shows that these filaments
have kinetic temperatures greater than 20~K and densities
smaller than $1000$~cm$^{-3}$. They have a well defined
column density $N_{\rm CO}/\Delta v=3\times10^{15}\rm
cm^{-2} (km~s^{-1})^{-1}$ and $^{12}$CO(1-0) opacity
$\approx0.2$. Compared to the material detected in $^{13}$CO
($T_{kin}=8-9$~K and $n_{\rm H_2}\approx1-2\times10^3\rm
cm^{-3}$), these $^{12}$CO-filaments are thus warmer and
more tenuous.

\subsection{Magnetic fields}

The analysis of the structure in the $^{12}$CO and $^{13}$CO
with automatic procedures \citep{stutzki1990} shows that the
emission from both isotopologues arise in elongated
filamentary structures, whose orientation is not random. The
distribution of their position angle is peaked around
P.A.=$105\pm40$\deg. Interestingly, the plane of the sky
projection of the magnetic fields orientation, measured 4~pc
North, has P.A.=$108\pm19$\deg\ \citep{heiles2000}.

\section{Dissipative structures of interstellar turbulence}


An attested deviation to the early Kolmogorov's turbulence
theory is the fact that dissipation is intermittent, in
space and time: the fraction of the volume and the timescale
over which dissipation is active decreases with spatial
scales \citep{ll,k62,frisch1995}. While the turbulent velocity
follows a Gaussian distribution, intermittency makes its
increments over a distance $l$ to depart from a Gaussian:
tails show up because large increments are more numerous
than in a Gaussian distribution. This non-Gaussianity is
more pronounced when decreasing $l$. As a consequence of
intermittency, the velocity shear ($\partial_{i}v_j$)
reaches large values at small scale, that manifest themselves  
as bursts of viscous dissipation
($\propto\sigma^2=\frac{1}{2}\sum_{ij}
(\partial_{i}v_j+\partial_{j}v_i)^2$, with $\mathbf{v}$ the
 fluid velocity). These violent events are associated
to the non-Gaussian tails of the distribution of the
velocity shear.


\subsection{Intermittency in the interstellar molecular gas}

Interstellar turbulence can only be partially probed: we
measure the gas velocity projected along the line of sight
(LOS) and the position projected in the plane of the sky
(POS). All observables are LOS integrals: they result from a
complex combination of radiative transfer in lines,
chemistry, density and velocity structures of the gas.
\cite{lis1996} have shown that in the case of optically thin
lines, the increments of line centroid velocities (CVIs)
measured between two different LOS trace a quantity related
to the LOS average of the
POS projection of the velocity shears.  It is therefore possible,
from the CVIs statistics computed in a map of spectra, to
approach that of the velocity shear, 
and find the subset of space
where the departures from a Gaussian distribution occur.



We have computed maps of CVIs for different lags, in the
$^{12}$CO(1-0) data, following
\cite{pety2003}. Fig.~\ref{fig:cvi} (left) shows their
probability density function for two lags $l=18$ pixels (the
largest lag with a significant number of pairs of points)
and $l=3$ pixels (the smallest between independent
points). The $l=18$ distribution (blue) is very close to a
Gaussian (deviations are due to large scale
gradients). However, the $l=3$ pixel one (red) has non-Gaussian wings
that show up for large velocity increments. We show below and in Falgarone 
et~al. (this volume) 
why these wings may be 
ascribed to the space-time intermittency of turbulence 
dissipation.


\subsection{Filaments of extreme velocity shear}

\begin{figure}[!t]
  \begin{center}
	\includegraphics[width=0.40\textwidth,angle=00]{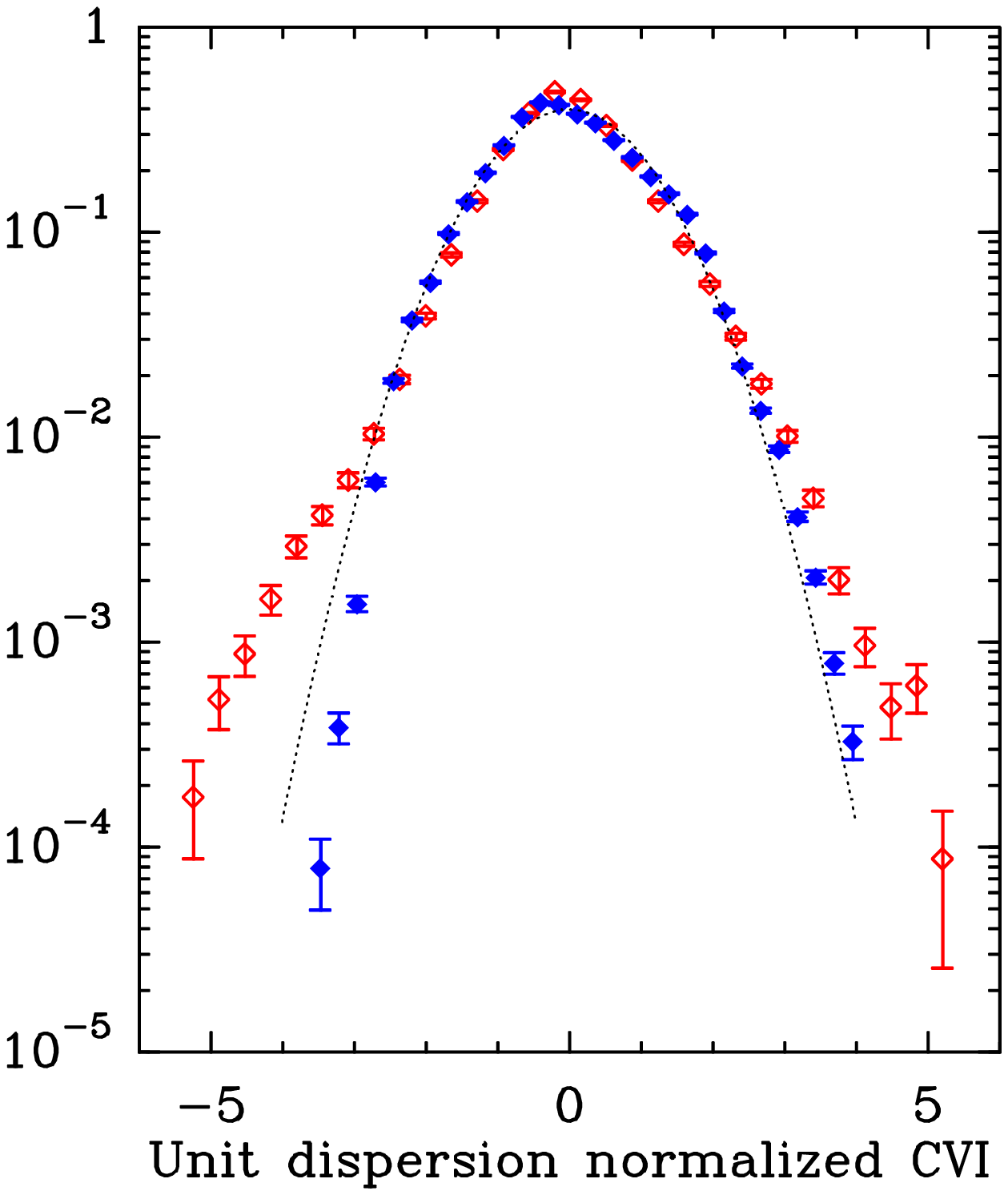}\hfill%
	\includegraphics[width=0.40\textwidth,angle=0]{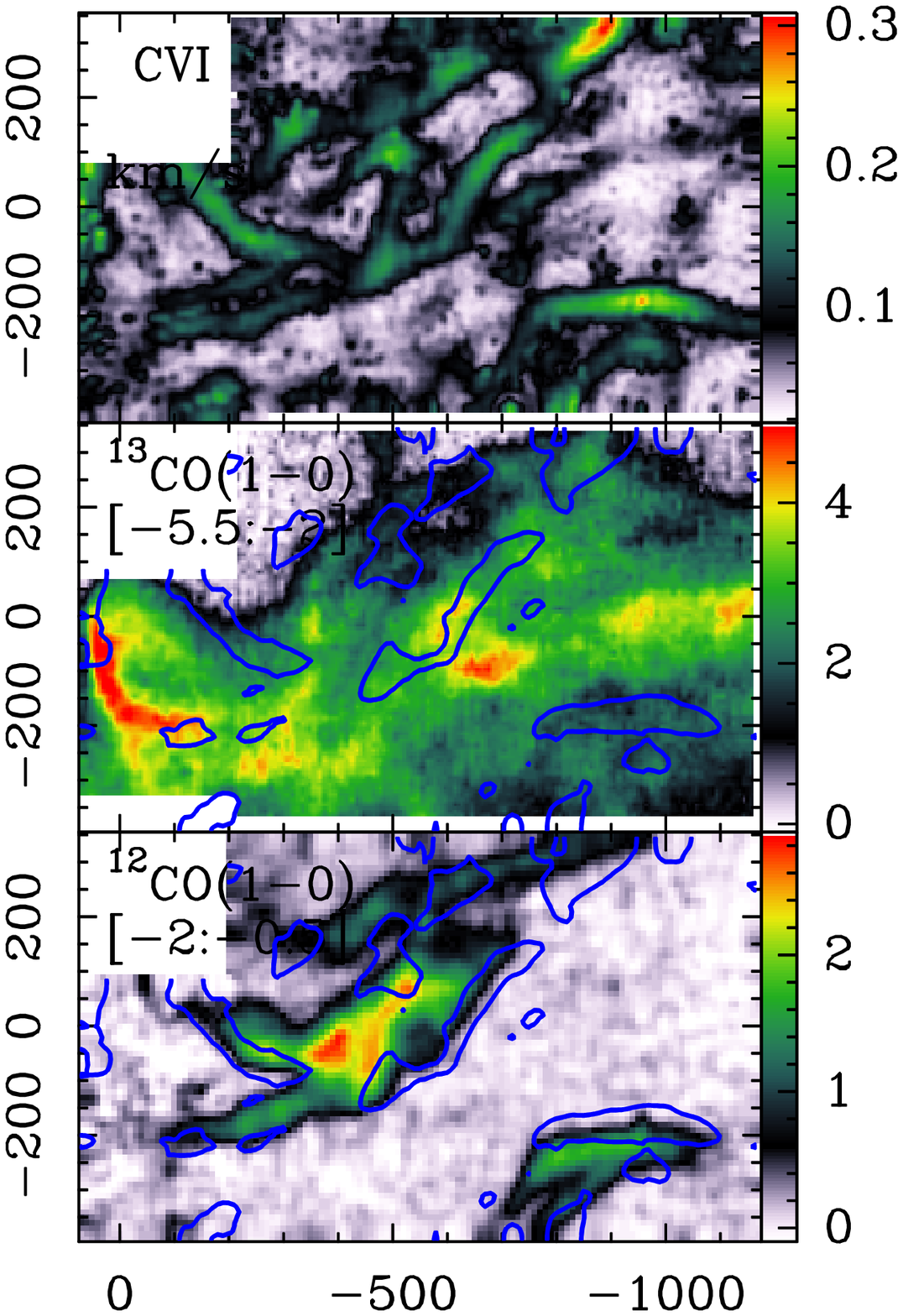}
	\caption{\textit{Left panel}: Probability density
	  function (PDF) of the centroid velocity increments,
	  normalized to unit dispersion (in blue, for a $l=18$
	  pixels, in red $l=3$ pixels).  \emph{Right panels}:
	  from top to bottom, \textit{i)} spatial distribution
	  of the centroid velocity increments in the
	  $^{12}$CO(1-0) emission (~km~s$^{-1}$), \textit{ii)}
	  integrated emission of the $^{13}$CO(1-0),
	  \textit{iii)} $^{12}$CO emission in the range
	  $[-2:-0.5]$~km~s$^{-1}$ (line-wing emission). 
          The 0.12~km~s$^{-1}$ CVI
	  contours are indicated in blue.}
	\label{fig:cvi}
  \end{center}
\end{figure}

Fig.~\ref{fig:cvi} shows the map of CVIs, computed for a
3-pixel lag, on the $^{12}$CO(1-0) data (top right panel). 
The largest CVIs, those
points populating the wings in the PDF (left), 
form elongated structures
of aspect ratio as high as 10. These structures are
poorly correlated with the $^{13}$CO emission (middle right
panel). On the contrary, their spatial correlation
with the optically thin $^{12}$CO emission (bottom right
panel) is striking. The sites of largest velocity shear
are therefore associated with gas warmer than average (see Sect. 2.1).

\subsection{Coherence from small to large scales}

\begin{figure}
  \begin{center}
	\includegraphics[width=0.43\textwidth]{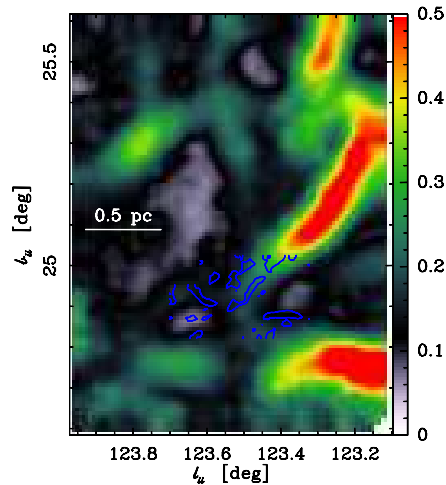}
	\caption{CVI map calculated on the large-scale Kosma
	  data \citep{bensch2001} with the contours from the top
	  left panel of Fig.~\ref{fig:cvi}.}
	\label{fig:kosma}
  \end{center}
\end{figure}

We applied the same procedure to large scale data obtained
with coarser spatial resolution \citep{bensch2001}. The
resulting CVIs (Fig.~\ref{fig:kosma}) computed for a small
lag (3 pixels) again show filaments, extending over 1 parsec
or more. Interestingly, the longest shear-filament
extends the small one found in
the 30m data set. The south-western filament at large scale
might also be connected to the small one found in our maps.
It is however not possible to ascertain whether these large
scale shear filaments also correspond to warm and tenuous
molecular gas.

These two datasets, treated in similar ways, show that the
regions of largest velocity shear are high-aspect ratio
($>20$) filaments coherent over the parsec scale , despite
the fact that the molecular flow is turbulent.

\section{Discussion and summary}

The computation of the CVIs, at both small and large scales,
show that \emph{i)} the PDF of CVI is not
Gaussian for small lags, a possible signature of
intermittency, \emph{ii)} the points populating the
non-Gaussian wings of the PDF are distributed into
filaments, \emph{iii)} these filaments are associated with
gas warmer and more tenuous than the bulk of 
molecular gas, and \emph{iv)} these filaments remain
coherent over more than 1~pc. In this particular field, we
also found that the filaments are 
preferentially aligned along the direction 
of the magnetic fields. This suggests that
magnetic fields may be strong enough to imprint their
configuration on the gas structure.

The spatial correlation between the warm optically thin
$^{12}$CO(1-0) filaments and the regions of large velocity
shear suggests that these structures are sites of burst of
viscous dissipation which may be the dominant heating source. Further
non-equilibrium chemical modelling of HCO$^+$(1-0)
observations in these structures does indeed stress the
importance of viscous and ion-neutral--drift heating (see
Falgarone \emph{et al}, this conference).



\begin{thebibliography}{}
\expandafter\ifx\csname natexlab\endcsname\relax\def\natexlab#1{#1}\fi

\bibitem[{{Abergel} {et~al.}(1994){Abergel}, {Boulanger}, {Mizuno}, \&
  {Fukui}}]{abergel1994}
{Abergel}, A., {Boulanger}, F., {Mizuno}, A., \& {Fukui}, Y. 1994, \apjl, 423,
  L59
\bibitem[{{Bensch} {et~al.}(2001){Bensch}, {Stutzki}, \&
  {Ossenkopf}}]{bensch2001}
{Bensch}, F., {Stutzki}, J., \& {Ossenkopf}, V. 2001, \aap, 366, 636

\bibitem[{{Cambr{\' e}sy}(1999)}]{cambresy1999}
{Cambr{\' e}sy}, L. 1999, \aap, 345, 965

\bibitem[{{Crutcher}(1999)}]{crutcher1999}
{Crutcher}, R.~M. 1999, \apj, 520, 706

\bibitem[{{Falgarone} {et~al.}(1998){Falgarone}, {Panis}, {Heithausen},
  {Perault}, {Stutzki}, {Puget}, \& {Bensch}}]{falgarone1998kp}
{Falgarone}, E., {Panis}, J.-F., {Heithausen}, A., {et~al.} 1998, \aap, 331,
  669

\bibitem[{{Falgarone} \& {Phillips}(1996)}]{falgarone1996}
{Falgarone}, E. \& {Phillips}, T.~G. 1996, \apj, 472, 191

\bibitem[{{Frisch}(1995)}]{frisch1995}
{Frisch}, U. 1995, {Turbulence. The legacy of A.N. Kolmogorov} (Cambridge
  Univ. Press)

\bibitem[{{Gerin} {et~al.}(1997){Gerin}, {Falgarone}, {Joulain}, {Kopp}, {Le
  Bourlot}, {Pineau des Forets}, {Roueff}, \& {Schilke}}]{gerin1997}
{Gerin}, M., {Falgarone}, E., {Joulain}, K., {et~al.} 1997, \aap, 318, 579

\bibitem[{{Heiles}(2000)}]{heiles2000}
{Heiles}, C. 2000, \aj, 119, 923

\bibitem[{{Heithausen}(2002)}]{heithausen2002}
{Heithausen}, A. 2002, \aap, 393, L41

\bibitem[{{Heithausen}(2006)}]{heithausen2006}
{Heithausen}, A. 2006, \aap, 450, 193

\bibitem[{{Hily-Blant} \&
{Falgarone}(2006)}]{hilyblant2006ii} {Hily-Blant}, P. \&
{Falgarone}, E. 2006, accepted by \aap

\bibitem[{Kolmogorov}(1962)]{k62} Kolmogorov, A.N. 1962 JFM 13, 82

 
\bibitem[{Landau} \& {Lifchitz}(1959)]{ll} Landau L.D. \& Lifchitz E.M. 1959, Fluid Mechanics, Addison-Wesley 

\bibitem[{{Lis} {et~al.}(1996){Lis}, {Pety}, {Phillips}, \&
  {Falgarone}}]{lis1996}
{Lis}, D.~C., {Pety}, J., {Phillips}, T.~G., \& {Falgarone}, E. 1996, \apj,
  463, 623

\bibitem[{{Miville-Desch{\^e}nes} \& {Lagache}(2005)}]{miville2005}
{Miville-Desch{\^e}nes}, M.-A. \& {Lagache}, G. 2005, \apjs, 157, 302

\bibitem[{{Padoan} {et~al.}(2001){Padoan}, {Juvela}, {Goodman}, \&
  {Nordlund}}]{padoan2001}
{Padoan}, P., {Juvela}, M., {Goodman}, A.~A., \& {Nordlund}, {\AA}. 2001, \apj,
  553, 227

\bibitem[{{Pan} {et~al.}(2001){Pan}, {Federman}, \& {Welty}}]{pan2001}
{Pan}, K., {Federman}, S.~R., \& {Welty}, D.~E. 2001, \apjl, 558, L105


\bibitem[{{Pety} \& {Falgarone}(2003)}]{pety2003}
{Pety}, J. \& {Falgarone}, E. 2003, \aap, 412, 417

\bibitem[{{Stutzki} \& {G{\" u}sten}(1990)}]{stutzki1990}
{Stutzki}, J. \& {G{\" u}sten}, R. 1990, \apj, 356, 513

\end{thebibliography}

\end{document}